\documentclass[12pt]{article}

\pdfoutput=1 

\usepackage{graphicx}
\usepackage{amssymb,amsmath,euscript,array}
\usepackage{subfigure}
\usepackage{color}
\usepackage{cite}
\usepackage{footnote}
\usepackage{caption}
\usepackage{hyperref}

\makeatletter
\@addtoreset{equation}{section}
\makeatother

\definecolor{holger}{rgb}{0,0.5,0.7}
\definecolor{edit}{rgb}{1,0,0}

\definecolor{durbeer}{rgb}{1,0,0}

\definecolor{durbeer2}{rgb}{0.8,0,0.5}

\newcommand{\hwpp}{\textsf{Herwig++}}

\newcounter{multieqs}

\newcommand{\gev}{{\rm GeV}~}
\newcommand{\tev}{{\rm TeV}~}

\def\bd{\begin{document}}
\def\ed{\end{document}}
\def\nn{\nonumber}
\def\bea{\begin{eqnarray}}
\def\eea{\end{eqnarray}}
\let\bm=\bibitem
\let\la=\label

\begin{document}

\hfill{IPPP/11/14; DCPT/11/28}\\[-0.9cm]

\vspace{25pt}

\begin{center}

{\Large \bf  New Constraints on Gauge Mediation and Beyond\\[1.5ex] 
from 
LHC SUSY
Searches at 
7 TeV
}

\vspace{35pt}

{\large \bf  Matthew J. Dolan, David Grellscheid, Joerg Jaeckel, \\[1.5ex] Valentin V. Khoze and Peter Richardson}

\medskip

{\small \em
{Institute for Particle Physics Phenomenology, Department of Physics, Durham
University,\\ Durham DH1 3LE, United Kingdom}}

\vspace{10pt}

{\sffamily \tt
}

\vspace{30pt}
\end{center}

\begin{abstract}
\noindent
The first results from the LHC on jets plus missing energy
provide powerful new data to test SUSY models. Initial theoretical interpretations of these data
have concentrated on gravity mediation, usually the CMSSM and its variations. In this paper we
confront a large class of gauge mediation models with these new data. 
More precisely we consider models of pure general gauge mediation (pure GGM) and confront them with the recent experimental results of the ATLAS collaboration.
We use \hwpp~and \textsf{RIVET}, incorporating the full set of experimental cuts, to calculate
the signal rates and compare them to the data.
Although based on only 35pb$^{-1}$ of integrated luminosity, we show that these new data probe and exclude a portion
of previously allowed parameter space of GGM.\\
In addition we investigate the viability of standard SUSY benchmark points, including
the Snowmass, CMS and ATLAS sets which encompass other mediation scenarios such as gravity,
anomaly and gaugino mediation.
\end{abstract}

\setcounter{page}{0}
\thispagestyle{empty}
\newpage

\section{Introduction}

Recently the CMS and ATLAS collaborations conducted a first series of searches for supersymmetry in 7 TeV proton-proton collisions
at the LHC looking for squarks and gluinos in final states containing jets and missing 
energy~\cite{Khachatryan:2011tk,daCosta:2011hh,daCosta:2011qk,Chatrchyan:2011wc}.
These analyses were based on $35\, {\rm pb}^{-1}$ of data taken in 2010.
As no excess above the Standard Model expectations was observed in these experiments,
their results set limits on the sparticle production and consequentially provide stringent new constraints
on the allowed regions of  parameter spaces in SUSY models. So far, most of the theoretical analysis of these initial LHC searches
has concentrated on constraining the CMSSM and related models \cite{Allanach:2011ut,Buchmueller:2011aa,Bechtle:2011dm,Allanach:2011wi,Akula:2011zq,Akula:2011dd} which are expected to arise from gravity mediation scenarios.

In this paper we will assess the impact of these searches for gauge mediated SUSY breaking models.
The most stringent constraints on the CMSSM\footnote{In this paper we consider gauge mediation models where the next-to-lightest supersymmetric particles~(NLSP) are stable on collider timescales. Limits on models where the NLSP decays promptly into photons
and gravitinos are presented in \cite{Chatrchyan:2011wc}.}
currently come from the ATLAS zero-lepton searches \cite{daCosta:2011qk}. Here we will concentrate upon these,
and apply them to a concise class
of General Gauge Mediation (GGM) known as pure GGM \cite{Abel:2009ve,Abel:2010vba}.
The results of this analysis are summarized in Fig.~\ref{nessie}.
We will also comment on constraints for ordinary gauge mediation in Section~\ref{sect:ogm}.

Going beyond gauge mediation, we use the LHC data to test the viability of the Snowmass sps~\cite{Allanach:2002nj}, the ATLAS SU~\cite{Aad:2009wy} and the CMS LM~\cite{Ball:2007zza} benchmark points, which represent a large variety of mediation scenarios (cf. Tab.~\ref{tab:benchmark}). For comparison we combine and interpret constraints on the CMSSM, pure GGM and the benchmark points in a model independent way in terms of physical squark and gluino masses, in Figs.~\ref{masses1}-\ref{masses3}.

\section{Gauge Mediation}
Theories with gauge mediated supersymmetry breaking
provide a particularly simple and compelling set-up
for addressing theory and phenomenology beyond the Standard Model, see \cite{Giudice:1998bp} for a review. 
On the theoretical side, gauge
mediation provides an advantage compared to other SUSY-breaking mechanisms (such as gravity mediation)
due to its automatic avoidance of unsuppressed flavour changing interactions\footnote{On the other hand, gauge mediation does not
provide any straightforward WIMP dark matter candidates. Contrary to gravity mediation, the lightest neutralino in gauge mediation
will always ultimately decay to the gravitino, which is the LSP. This rules out neutralino candidates for dark matter
in gauge mediation, though not the possibility of gravitino dark matter.}.
Over the last few years there has been a surge of interest in gauge mediation which has led to a significant extension and
generalisation of its original realisation.
The GGM framework, first introduced in \cite{Meade:2008wd}, is suitable for
unifying quite general models of gauge mediation in a model-independent way. 
It requires that supersymmetry breaking
is communicated to the Standard Model (MSSM) sector through gauge
interactions at the messenger scale.

\subsection{Pure general gauge mediation}
The GGM formulation (or more precisely pure GGM) is based on the requirement that SUSY-breaking effects
in the MSSM should disappear in the limit of vanishing Standard Model gauge couplings. The resulting description does
not require precise knowledge of any specific underlying models, which
can be weakly or strongly coupled, with explicit messenger sectors or direct mediation, or any combination of the above.

A detailed study of the
phenomenology of pure GGM models and their parameter spaces was presented recently in \cite{Abel:2009ve,Abel:2010vba}.
As alluded to above, in pure GGM we have no direct couplings
of the SUSY-breaking sector to the Higgs sector, and therefore the soft parameter
$B_{\mu}$ is approximately zero at the
messenger scale. From this starting point at the
high scale $M_{{\rm mess}}$ a small but viable value of $B_\mu$ is
generated radiatively at the electroweak scale \cite{Rattazzi:1996fb,Babu:1996jf}.
Electroweak symmetry breaking then determines the
values of $\tan\beta$ and $\mu$.
Since $B_\mu$ is small, $\tan\beta$ is generally
large (between $20$ and $70$).

The main free parameters of pure GGM models are the gaugino and scalar masses as well as
the messenger scale~\cite{Jaeckel:2011ma,Jaeckel:2011qj}. In models with messenger fields transforming in complete and unsplit GUT multiplets, there is a single effective
scale $\Lambda_G$ for the gaugino masses and a single scale $\Lambda_S$ for the
scalars~\cite{Jaeckel:2011qj}\footnote{More generally, if one does not require unification, there are six distinct $\Lambda$ scales in GGM:
$\Lambda_{G,r}$, $\Lambda_{S,r}$ with $r=1,2,3$ in GGM~\cite{Meade:2008wd}.}.
Generating $\Lambda_{G}$ requires both R-symmetry and supersymmetry breaking while $\Lambda_{S}$ is affected only by supersymmetry breaking. For this reason $\Lambda_{G}$ and $\Lambda_{S}$ are two a priori distinct scales in GGM. In the simplest scenario, ordinary gauge mediation (on which we will comment in the next subsection), one can 
nevertheless identify these two scales, $\Lambda_{G}\simeq \Lambda_{S}$.

In GGM the soft supersymmetry breaking gaugino masses at the messenger scale $M_{{\rm mess}}$ are given by
\begin{equation}
\label{gauginosoft}
M_{\tilde{\lambda}_i}(M_{{\rm mess}}) =\, k_i \,\frac{\alpha_i(M_{{\rm mess}})}{4\pi}\,\Lambda_G
\end{equation}
where $k_i = (5/3,1,1)$, $k_i\alpha_i$ (no sum)
are equal at the GUT scale and $\alpha_i$ are the gauge coupling constants.
Similarly, the scalar mass squareds are
\begin{equation}
\label{scalarsoft}
m_{\tilde{f}}^2 (M_{{\rm mess}}) =\, 2 \sum_{i=1}^3 C_i k_i \,\frac{\alpha_i^2(M_{{\rm mess}})}{(4\pi)^2}\, \Lambda_S^2
\end{equation}
where the $C_i$ are the quadratic Casimir operators of the gauge groups.

In summary, the value of the high scale $M_{{\rm mess}}$, together with $\Lambda_G$ and $\Lambda_S$ appearing in Eqs.~\eqref{gauginosoft}-\eqref{scalarsoft}
at  $M_{{\rm mess}}$ characterise a point in the pure GGM
parameter space. In this sense pure GGM is the gauge mediation analogue of the CMSSM and mSUGRA models
with $\Lambda_G$ and $\Lambda_S$ playing a role similar to the parameters $m_{\frac12}$ and
$m_0$ in those models. However these gravity mediated simple realisations are physically quite distinct from the
gauge mediated pure GGM framework we analyse here. The main differences in gauge mediation include:
\begin{itemize}
\item the gravitino is always the LSP and the NLSP can be long lived and is not 
necessarily neutral;
\item at the high scale the sfermion masses are not identical, in particular the left- and right-handed sfermions
have different masses as can be seen from Eq.\,\eqref{scalarsoft};
\item $M_{\rm mess}$ is a parameter of the model which is typically much lower than the GUT scale.
\end{itemize}

\begin{figure}[t]
\begin{center}
\includegraphics*[viewport= 130 70 440 370, width=0.6\textwidth]{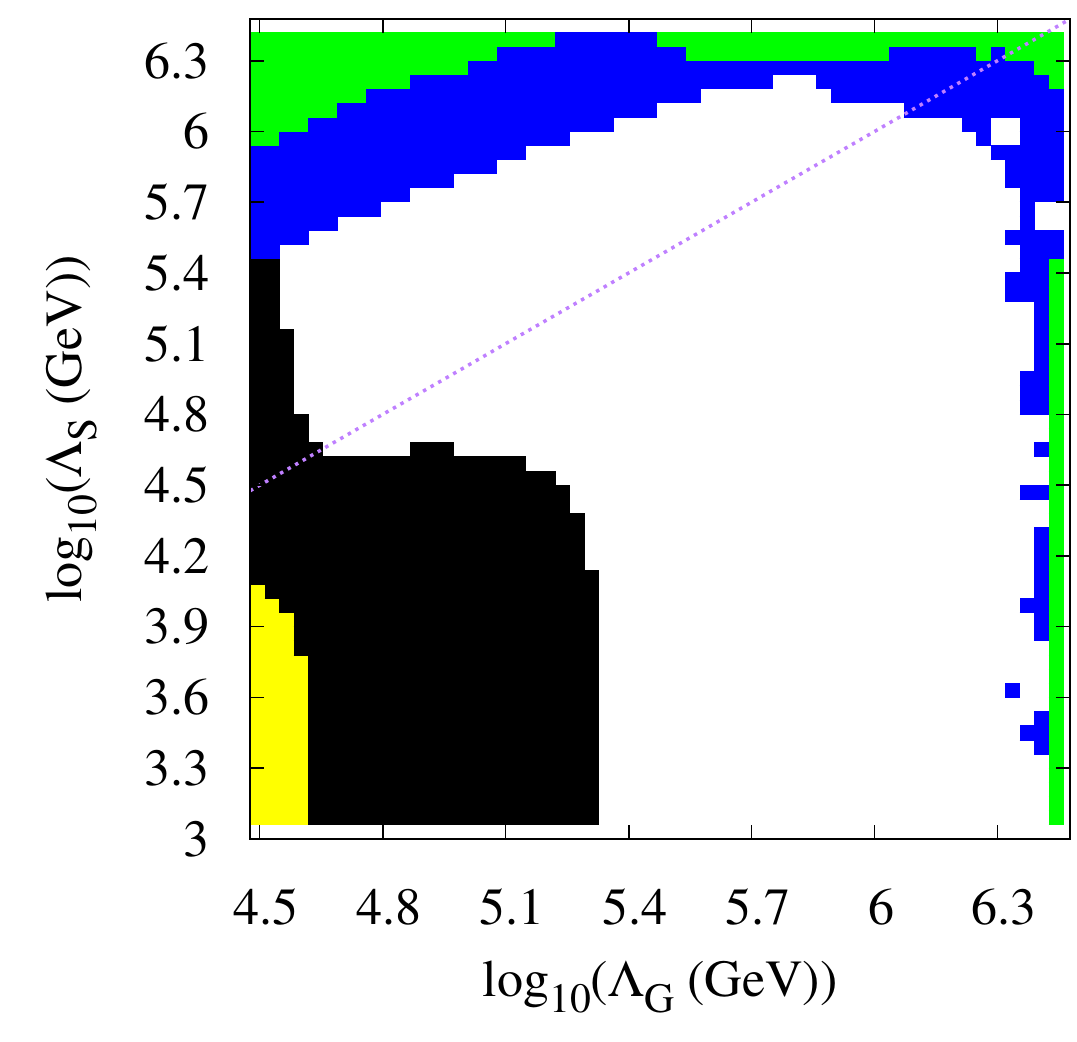}
\caption{
Pure GGM parameter space for intermediate messenger scales,  $M_{{\rm mess}} = 10^{10}$~GeV. The
dominant constraints excluding various areas around the allowed (white) region are indicated as follows:
points in the black region violate the pre-LHC direct search limits, while
yellow area is excluded by the presence of tachyons in the spectrum.
In the blue region SoftSUSY has not converged and in the green
region a coupling reaches a Landau pole during RG evolution.
Ordinary gauge mediation lives on the dotted line.}
\label{phenoland}
\end{center}
\end{figure}

Outside the confines of \emph{ordinary} gauge mediation, where the $\Lambda_G\simeq \Lambda_S$,
the pure GGM parameter space is populated by many models that
predict different values of the ratio of gaugino to scalar masses, $\Lambda_G/\Lambda_S$. This parameter space was investigated
in Refs.~\cite{Abel:2009ve,Abel:2010vba} from which we adopt
Figure~\ref{phenoland}. This figure shows the allowed parameter space of pure GGM in the $(\Lambda_G,\Lambda_S)$~plane (before the
new LHC constraints are imposed) for a fixed value of $ M_{{\rm mess}} = 10^{10}$~GeV. We used a modified version of
\texttt{SoftSUSY}~\cite{Allanach:2001kg}, which takes $B_{\mu}=0$ as an input and predicts $\tan\beta$
using the electroweak symmetry breaking conditions.
Direct experimental searches from the Tevatron
and LEP (see~\cite{Abel:2010vba} for more detail) exclude
the black region in Fig.~\ref{phenoland} on the left of the Nessie-shaped pure GGM parameter space, not surprisingly it effectively cuts off the lower
values of gaugino and scalar masses. Other boundaries of the parameter space arise from requiring that there are no
tachyons and no Landau poles, and that SoftSUSY has not encountered convergence problems during the RG evolution between the high
and the low scales.

\begin{figure}
\begin{center}
\vskip -1cm
\subfigure{
\includegraphics*[viewport= 142 75 500 385,width=6.7cm]{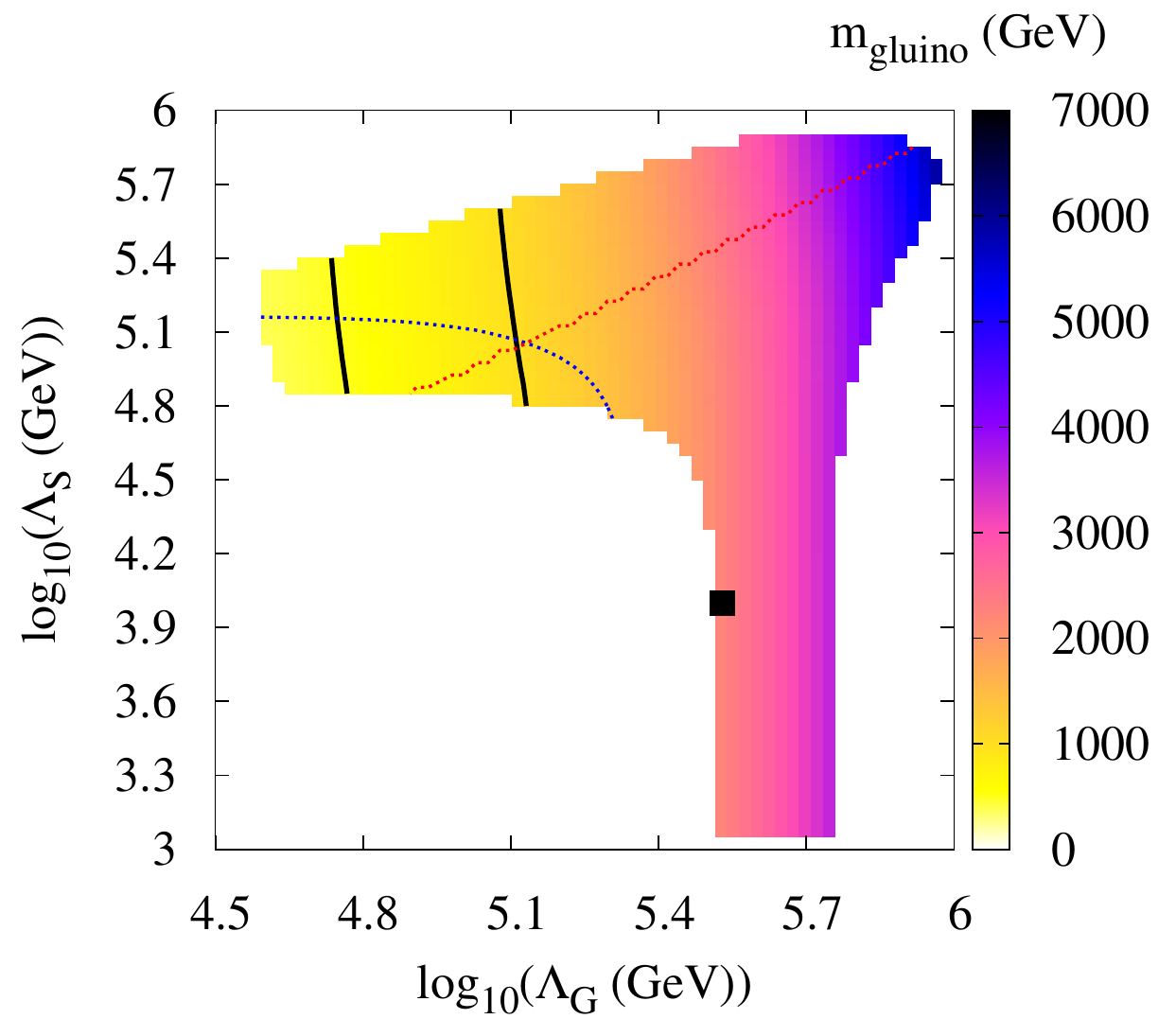}
}
\subfigure{
\includegraphics[width=6.8cm]{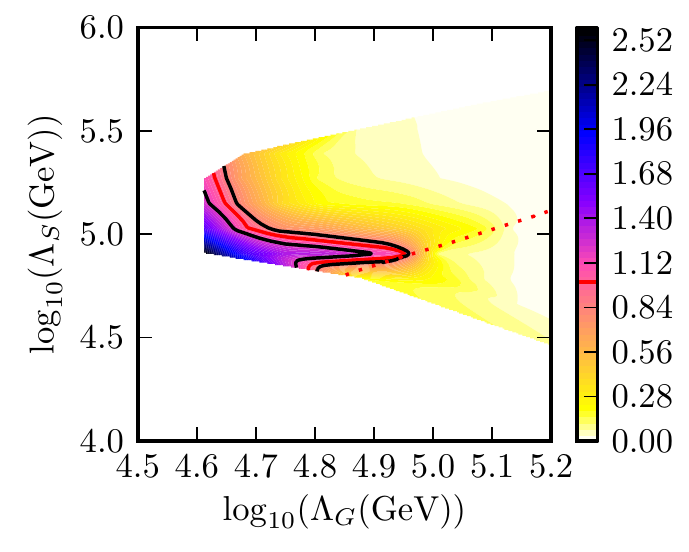}
}
\subfigure{
\includegraphics*[viewport= 142 75 500 385,width=6.7cm]{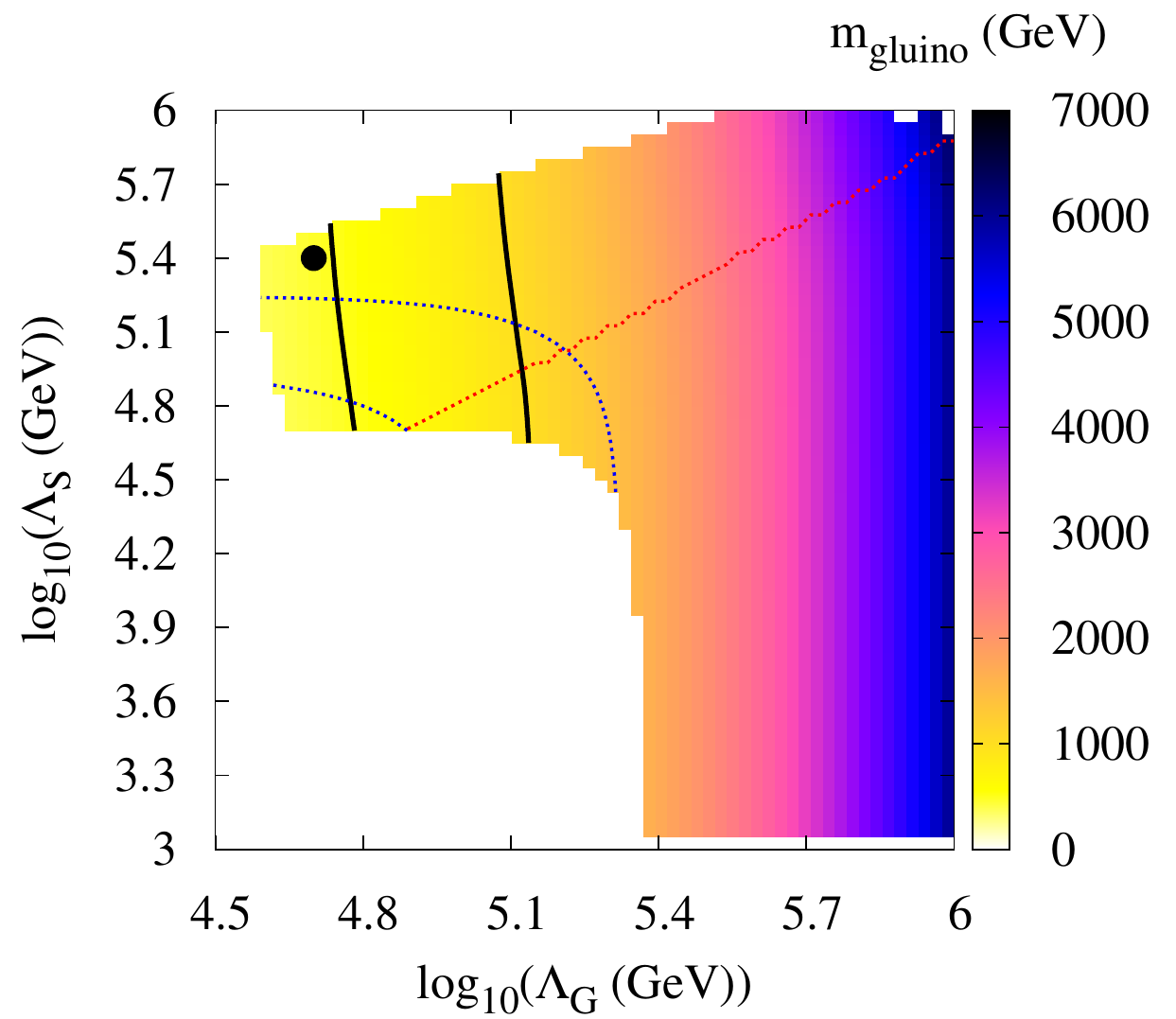}
}
\subfigure{
\includegraphics[width=6.8cm]{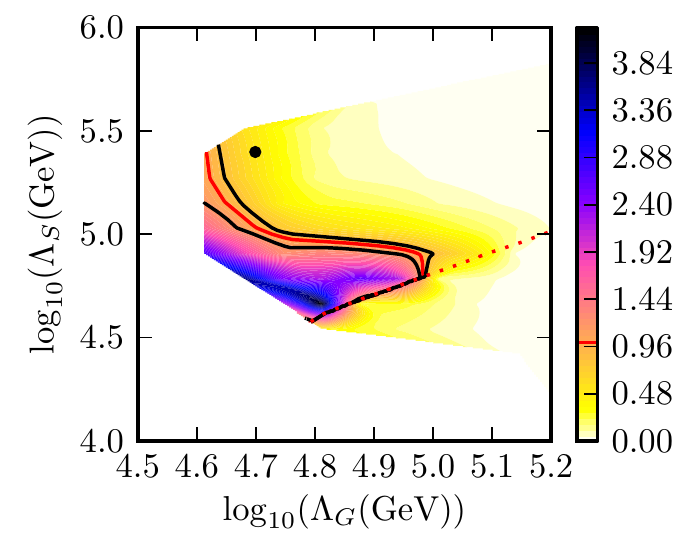}
}
\subfigure{
\includegraphics*[viewport= 142 75 500 385,width=6.7cm]{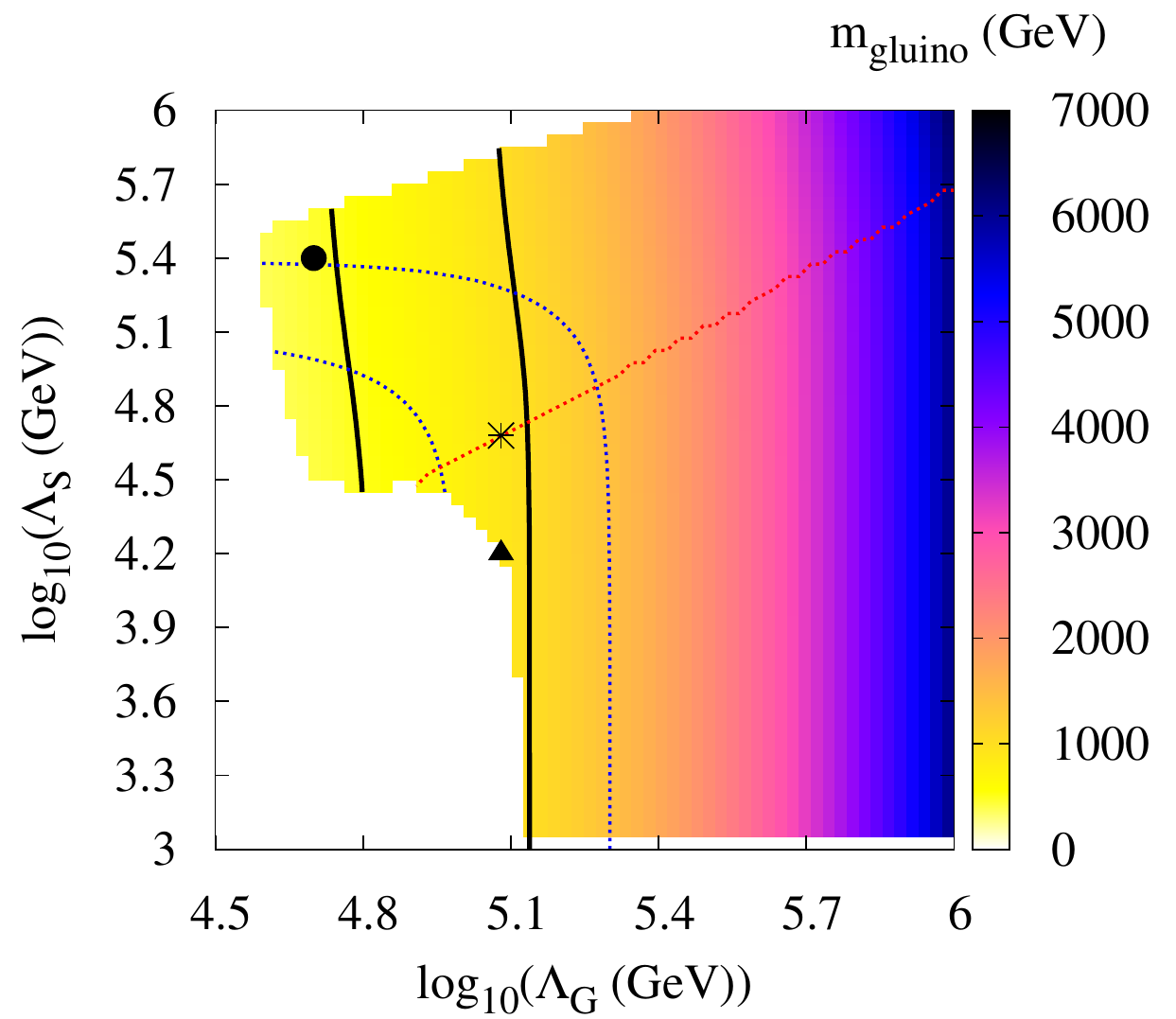}
}
\subfigure{
\includegraphics[width=6.8cm]{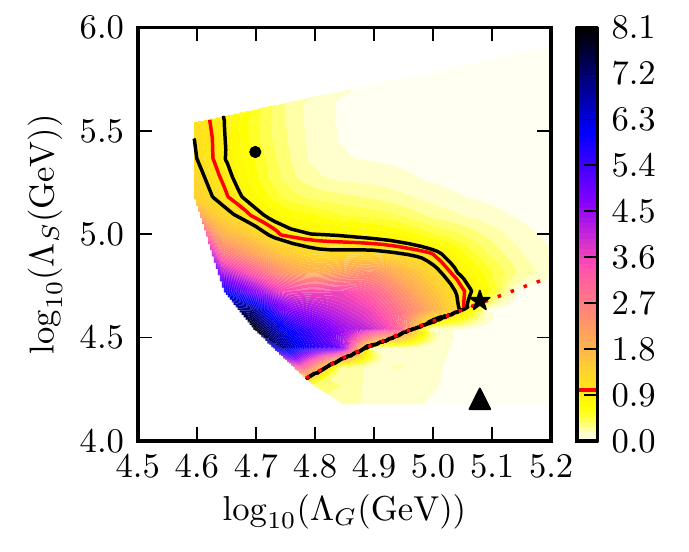}
}
\caption{ 
The left panels show the pure GGM parameter space in terms of $\Lambda_G$,~$\Lambda_S$ defined in Eqs.~\eqref{gauginosoft} and \eqref{scalarsoft}. 
From top to bottom we have $M_{{\rm mess}}=10^{8}\,{\rm GeV},\,\, 10^{10}\,{\rm GeV}$ and $10^{14}\,{\rm GeV}$.
Stop mass contours (500~\gev and 1~\tev) are indicated as dotted lines, and the 500~\gev and 1~\tev  gluino lines are solid.
The NLSP is neutralino above the diagonal red line and stau below.
The panels on the right show 95\% exclusion contours derived from the ATLAS search as red lines, and the black lines indicate uncertainties due to scale variations in the NLO cross-section. The colour scale for the right panels shows the expected number of signal events normalised to the exclusion limit. The benchmark points discussed in~\cite{Abel:2010vb} are shown as
a dot (PGM1a middle panel, PGM1b bottom panel), triangle for PGM2, a star for PGM3 and finally a square for PGM4.}
\label{nessie}
\end{center}
\end{figure}

In gauge mediation models with explicit messengers one expects
the ratio $\Lambda_G/\Lambda_S$ to be close
to one (dotted line in Fig.~\ref{phenoland}), while for direct mediation models the gaugino masses are
often suppressed relative to the scalar
masses~\cite{Izawa:1997gs,Kitano:2006xg,Csaki:2006wi,Abel:2007jx,Abel:2007nr,Abel:2008gv,Komargodski:2009jf,Abel:2009ze}
(region above the dotted line in Fig.~\ref{phenoland}).
It is also possible to
achieve values $\Lambda_G/\Lambda_S>1$ by increasing the ``effective number of
messengers''~\cite{Kaplan:1999ac,Chacko:1999mi,Csaki:2001em,Cheung:2007es,McGarrie:2010qr,Green:2010ww} (region below the dotted line in Fig.~\ref{phenoland}).

The discovery potential for pure GGM models during the early stages of the LHC with \mbox{$\sqrt{s}=7\, {\rm TeV}$}
was addressed
in Ref.~\cite{Abel:2010vba}. The left panels in Fig.~\ref{nessie} show three scans of the parameter space of pure GGM 
(taken from~\cite{Abel:2010vba}), at different messenger scales, $M_{{\rm mess}}=10^{8}\,{\rm GeV},\,\,10^{10}\,{\rm GeV}$  and $10^{14}\,{\rm GeV}$.
The regions expected to be most sensitive to $pp$-scattering at \mbox{$\sqrt{s}=7\, {\rm TeV}$}, correspond to relatively light gluinos and/or relatively light squarks.
In each figure stop mass contours of 500~\gev and 1~\tev are indicated as dotted lines, and the 500~\gev and 1~\tev  gluino contours are indicated as solid lines\footnote{In the $M_{{\rm mess}}= 10^8$~\gev scenario the single dotted contour is for 1~\tev stop masses.}. Furthermore, the diagonal dotted red line corresponds to the boundary between neutralino and slepton NLSP.
The figures also contain the benchmark points introduced in~\cite{Abel:2010vba}. 

We now implement the new experimental constraints on SUSY searches obtained from the ATLAS data on final states
with jets, missing energy and no leptons. The details of our analysis of this data are explained in section~\ref{analysis}. Our results
are obtained from a Monte Carlo simulation of the signal events using 
\hwpp~\cite{Bahr:2008pv,Gieseke:2011na} and \textsf{RIVET}~\cite{Buckley:2010ar}, implementing all the experimental cuts imposed by 
ATLAS~\cite{daCosta:2011qk}. 
Combining the four signal regions defined by ATLAS~\cite{daCosta:2011qk} we obtain 
the constraints shown as the red lines in the right panels of Fig.~\ref{nessie}.
One can clearly see that already the relatively small data sample of 35~pb$^{-1}$ provides interesting new 
bounds on models of pure general gauge mediation.   

As one would expect the excluded regions correspond to relatively low gluino and squark masses. This will become
clearer in Fig.\,\ref{masses2} which shows the GGM exclusion region directly in terms of the squark 
and gluino masses~(rather than $\Lambda_S$ and $\Lambda_G$). It is interesting to note that there
nevertheless exists an allowed narrow wedge shaped region at low values of $\Lambda_S$, i.e. for relatively
low squarks masses. In this region the NLSP is a stau rather than a neutralino.
When the NLSP is the lightest stau, the ATLAS jets plus missing energy search
does not constrain the GGM model. As the NLSP is charged 
the only missing energy in the events comes from 
the production of neutrinos in the SUSY cascade decays. We would expect that the stau would
either be reconstructed as a muon, provided that its time delay reaching the muon chambers
is sufficiently short, or the event is rejected due to mismeasurement of the missing
transverse energy if its interaction in the muon chambers is not recorded. To be conservative we
do not consider these events in this paper. However we emphasize that this is a very interesting region of parameter space as it gives a smoking gun 
for gauge mediation. This is
worthy of further study using a similar approach to that in \cite{Aad:2011yf}.

We can compare the excluded regions with the previously identified best-fit regions from a global fit to low energy observables performed in~\cite{Abel:2009ve}. For both $M_{\rm{mess}}=10^{10}$~GeV and $10^{14}$~GeV the ATLAS search rules out some parameter space which was within the 95\% confidence limits of those fits~\footnote{For $M_{\rm{mess}}=10^8$~GeV such a fit has not yet been performed.}. For $M_{\rm{mess}}=10^{10}$~GeV a small region of the previous 68\% CL has now been ruled out at low $\Lambda_G$ and moderate $\Lambda_S$. The best-fit points, which lie deep in the stau NLSP region, are unaffected by the ATLAS results. This differs from recent fits of the CMSSM and related models\cite{Buchmueller:2011aa}, where the best-fit points were more strongly affected, although they still remain within the 95\% CL of the pre-LHC fits. The fact that in PGGM the best-fit points lie in the stau NLSP region further motivates a dedicated search for stau NLSPs.

We find that the lowest viable gluino mass which occurs is 380~GeV, a bound which is independent of the messenger scale and the identity of the NLSP. The lowest viable squark mass depends on the messenger and on the NLSP identity. For neutralino NLSP, we find that the lowest allowed squark mass is 735~GeV which happens when the gluino mass is approximately 500~GeV. Since we do not expect the ATLAS search to be sensitive to the stau NLSP region, the lowest permissible squark mass there is reduced to 490~GeV (where the squark mass is defined as the average mass of the first generation squarks). This can be read off from  Fig.~\ref{masses2}. 

\subsection{Ordinary gauge mediation}
\label{sect:ogm}
Models of ordinary gauge mediation live on the $\Lambda_{G}=\Lambda_{S}$ slice in GGM parameter space.
Here, we will also investigate the implications of the ATLAS data on these models.

In contrast to the discussion of the previous section here we will use a more traditional approach where $\tan\beta$ is treated as a
free parameter rather than a prediction.
This is achieved by deviating from the strict definition of gauge mediation and 
allowing for appropriate non-gauge couplings between messenger fields and the Higgs sector\footnote{Strictly speaking 
these extra couplings can also generate new contributions to the Higgs soft masses $m^{2}_{H_{u}}$ and $m^{2}_{H_{d}}$
but we will ignore these for simplicity.}. This can then generate an input value for $B_{\mu}$ at the high/messenger scale which is 
traded for $\tan\beta$ at the electroweak scale. As a result the essential parameter space is again 
three-dimensional, $\Lambda$, $M_{\rm {\rm mess}}$ and $\tan\beta$. In this scenario
the bound from the new LHC data for long lived NLSPs 
is fairly insensitive to the messenger mass with 
$\Lambda<72$\,TeV excluded by the ATLAS results~(for $\tan\beta=10$ although the bounds
are relatively independent of this parameter). This limit entirely arises
from the event selection C discussed in the next section and
corresponds to a squark mass 675\,GeV, a gluino mass of 590\,GeV and a NLSP~(lightest neutralino) mass of 94\,GeV.

\section{Implementation of the ATLAS event selection and analysis of the data}
\label{analysis}

  The ATLAS analysis~\cite{daCosta:2011qk} presents the number of observed events 
which pass four specific event selection criteria, together with the expected number of
Standard Model events. This can be used, together with an estimate of
the number of signal events passing the experimental cuts, to determine
whether a specific model is ruled out at the 95\% confidence level. Alternatively
the limits on the cross sections for non-SM processes given in~\cite{daCosta:2011qk} can be used.
Here we will do the latter as it already includes most of the statistical analysis.  

In order to compare the predictions of a particular BSM model with the ATLAS results
we therefore need to calculate the expected number of signal events passing the cuts in each of the four
regions~(A,B,C,D) defined in Table~1 of Ref.\,~\cite{daCosta:2011qk}.
These regions are designed to target light $\tilde{q}\tilde{q}$, heavy $\tilde{q}\tilde{q}$,
$\tilde{g}\tilde{g}$ and $\tilde{g}\tilde{q}$ production respectively, in the CMSSM. This is achieved by 
imposing different selection criteria on the number of jets~($\geq2$ in A and B and $\geq3$ in C and D)
as well as on the kinematics~($E^{\rm miss}_T$, $m_{\rm eff}$ and $m_{T2}$), see Ref.\,~\cite{daCosta:2011qk} for
details.

Each SUSY model is a point in the MSSM parameter space which is specified by the mass spectrum, SUSY couplings and mixing angles at the electroweak scale.
All these are contained in SLHA files produced by \textsf{SoftSUSY}~\cite{Allanach:2001kg} starting from the high-scale input from GGM or any other model.

Given the complexity of the signal processes
the calculation of the number of signal events 
is best achieved using a Monte Carlo event generator, in our case 
\textsf{Herwig++}~\cite{Bahr:2008pv,Gieseke:2011na}, to simulate the signal processes for
a given SUSY model. 
The experimental event selection can be implemented
using the \textsf{RIVET}~\cite{Buckley:2010ar} analysis 
framework\footnote{In addition we used the library based on the results of~\cite{Cheng:2008hk} to
calculate the $m_{T2}$ variable.} to analyse the hadronic final state
generated by the Monte Carlo simulation, without the need for a simulation of the detector
response.

\begin{figure}[t]
\begin{center}
\includegraphics[width=0.6\textwidth]{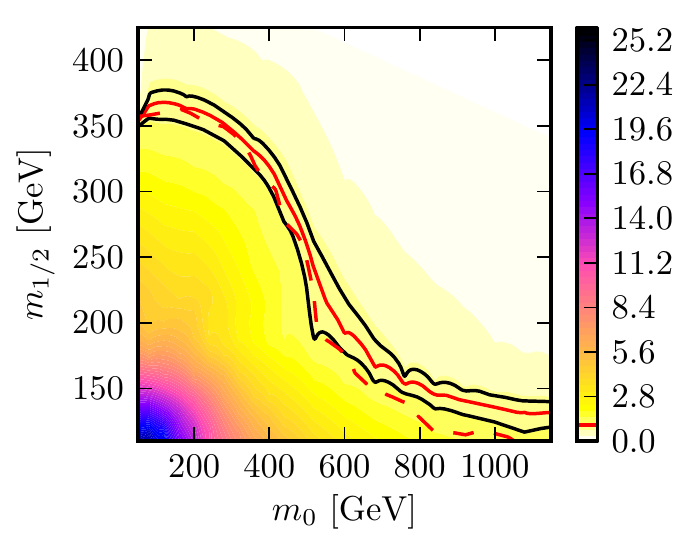}
\caption{95\% confidence level exclusion limit in the $(m_0,m_{\frac12})$~plane for
$\tan\beta=3$, $A_0=0$ and $\mu>0$ in the CMSSM. The solid red line is the result using
our signal simulations~(the solid black lines show the effect of
varying the factorization and renormalisation scales used to calculate
the next-to-leading order SUSY production cross sections by a factor of $\frac12$ and 2),
whereas the dashed red line is the limit obtained by ATLAS
in~\protect\cite{daCosta:2011qk}. The colour scale shows the expected number of signal events normalised to the exclusion limit.}
\label{fig:ATLAS-CMSSM}
\end{center}
\end{figure}

In principle this is sufficient to calculate the number of expected signal events
for any new physics model. However, for most new physics signals the matrix elements implemented
in general purpose Monte Carlo event generators are only accurate to leading order
in perturbative QCD. 
\textsf{Herwig++} was therefore used to simulate three sets of
supersymmetric particle production processes for 
each point in supersymmetric parameter space:
a) squark and gluino production, b) the production of an electroweak gaugino in association with 
a squark or gluino and c) the production of slepton and electroweak gaugino pairs. The fraction of events passing the experimental
cuts was then used together with the next-to-leading order cross section calculated using 
\textsf{Prospino}~\cite{Beenakker:1996ch,Beenakker:1999xh,Spira:2002rd,Plehn:2004rp}
to obtain the number of signal events passing the cuts for each of the four signal regions.

In order to check that our simulations and implementation of the experimental cuts was reliable 
we checked that we obtained good agreement with the numbers of events passing the cuts for
the CMSSM supersymmetric parameter points in $(m_0,m_{\frac12})$~plane for $\tan\beta=3$,
$A_0=0$ and $\mu>0$
supplied as supporting material~\cite{daCosta:2011qkWeb} with~\cite{daCosta:2011qk}.
The limit we obtain for the CMSSM 
using our simulations is in good agreement with that obtained by ATLAS\footnote{Except in the high $m_0$ region where
there is a significant scale uncertainty for the signal cross section, which we
have not included.},
as can be seen in Fig.\,\ref{fig:ATLAS-CMSSM}. These results use the NLO cross sections. As a cross check we have also
computed the bounds using the leading-order cross sections.
The resulting limits are somewhat lower but show  worse agreement with the ATLAS values.

\begin{figure}[!!t]
\begin{center}
\subfigure{
\includegraphics[width=7cm]{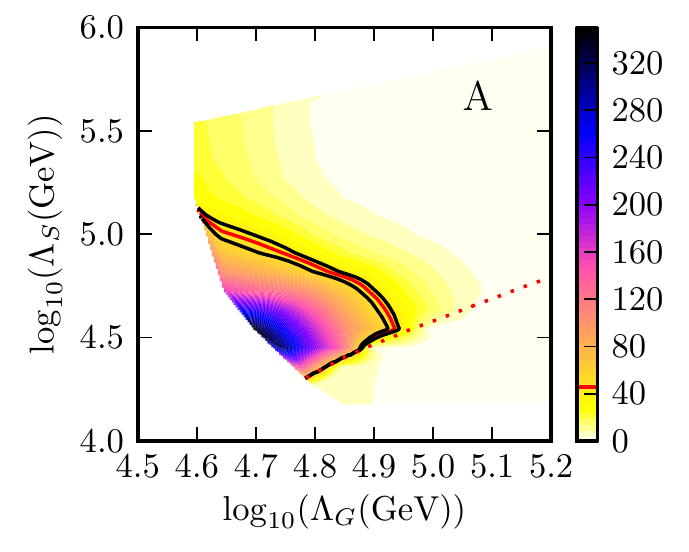}
}
\subfigure{
\includegraphics[width=7cm]{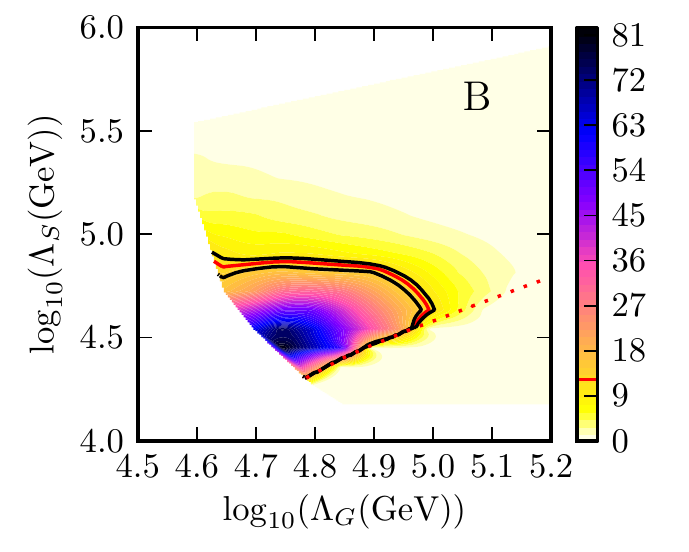}
}
\subfigure{
\includegraphics[width=7cm]{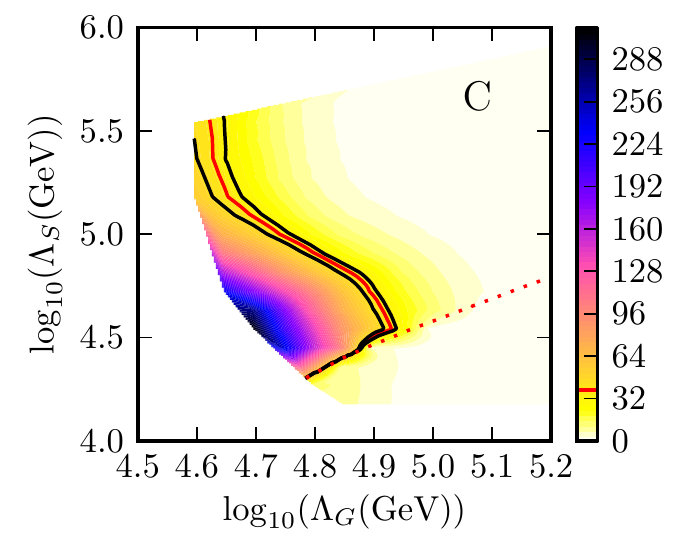}
}
\subfigure{
\includegraphics[width=7cm]{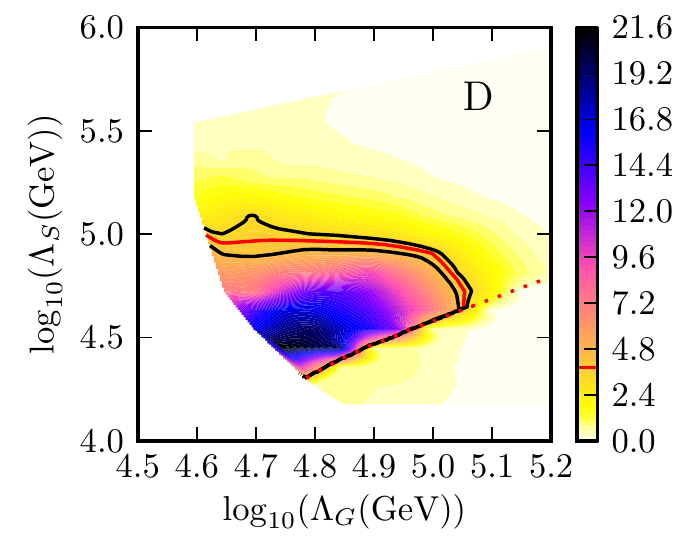}
}
\caption{Event rates of pure GGM ($M_{\rm mess}=10^{14}\,\,{\rm GeV}$) in the four signal regions corresponding to different sets of experimental cuts~\cite{daCosta:2011qk}. 
The shading gives the number of events predicted in our model, after
all cuts have been applied. The red line shows the exclusion contour. Below and to the right of the red dotted line the stau is the NLSP whereas above and to the left the NLSP
is the lightest neutralino. 
}
\label{ABCD}
\end{center}
\end{figure}

The number of events and the 95\% confidence level exclusion limit for pure GGM with 
$M_{\rm mess}=10^{14}\,\,{\rm GeV}$  is shown in Fig.\,\ref{ABCD} for each of the four ATLAS-defined regions. 
This exclusion limit is obtained from the maximum non-SM cross sections
of $1.3$, $0.35$, $1.1$ and $0.11$~pb, for regions A,B,C and D, respectively given
in \cite{daCosta:2011qk}.
As can be seen from Fig.\,\ref{ABCD} the strongest
limit is given by the C and D event selections with D giving the strongest limit at low
$\Lambda_S$ and C the strongest limit for high values of $\Lambda_S$. This is consistent with expectations from the design purpose of these regions, since at high $\Lambda_S$ the squark masses are high and the SUSY cross-section is dominated by $\tilde{g}\tilde{g}$ production, corresponding to ATLAS' region C At lower values of $\Lambda_S$ we are closer to squark-gluino mass degeneracy and so $\tilde{q}\tilde{g}$ production dominates, and region D provides the best search limits.
The combination of parameter space excluded by the various regions is given in the right panels of
Fig.\,\ref{nessie} for a range of messenger scales.

\section{Beyond gauge mediation: Analysis of full set of benchmark points}\label{sec:benchmark}
 
A number of benchmark points have been proposed for the study of supersymmetric
models at high energy colliders. There is a range of motivations for the 
selection of these points including: satisfying the current experimental constraints, providing the correct relic neutralino abundence to satisfy cosmological
limits and comparing the potentials of different experiments. A number
of these points, including the most studied sps1a, 
were designed to study the potential of measuring supersymmetric masses at the LHC
and therefore have relatively light mass spectra. As the LHC data now rule out
regions of the previously allowed parameter space,
nearly half of these points
are now excluded. In Table~\ref{tab:benchmark} we show effects of the ATLAS 0 leptons SUSY search on the Snowmass sps points, the ATLAS SU points, the CMS LM points and the PGM benchmarks. The columns of the table show the mediation scenario, the calculated cross-sections for each of the regions A,B,C and D, and finally whether the point is ruled out by the data and by which regions. Six of the ten sps points are ruled out by the ATLAS search, including sps1a. Also, four of the SU benchmarks are now excluded, and eight of the sixteen LM benchmarks. The PGM benchmarks proposed in~\cite{Abel:2010vba} remain allowed by the recent data. Since the proposed benchmarks are mostly at low masses, all the excluded points are ruled out at least by region D, if not by more regions.

\begin{table}
\begin{center}
{\footnotesize
\begin{tabular}{|c|c|c|c|c|c|c|}
\hline
Benchmark point& mediation scenario & \multicolumn{4}{|c|}{$\sigma/{\rm pb}$} & status\\\hline
& &A&B&C&D& ATLAS 35pb$^{-1}$\\\hline\hline
ATLAS Limits & & 1.3 & 0.35 & 1.1 & 0.11 & \\
\hline\hline
sps1a~\cite{Allanach:2002nj} & CMSSM & 2.031 & 0.933 & 1.731 & 0.418 & A,B,C,D\\\hline
sps1b~\cite{Allanach:2002nj} & CMSSM & 0.120 & 0.089 & 0.098 & 0.067 & allowed\\\hline
sps2~\cite{Allanach:2002nj}  & CMSSM & 0.674 & 0.388 & 0.584 & 0.243 & B,D\\\hline
sps3~\cite{Allanach:2002nj}  & CMSSM & 0.123 & 0.093 & 0.097 & 0.067 & allowed\\\hline
sps4~\cite{Allanach:2002nj}  & CMSSM & 0.334 & 0.199 & 0.309 & 0.144 & D      \\\hline
sps5~\cite{Allanach:2002nj}  & CMSSM & 0.606 & 0.328 & 0.541 & 0.190 & D      \\\hline
sps6~\cite{Allanach:2002nj}  & CMSSM~(non-universal $m_{\frac12}$) & 0.721 & 0.416 & 0.584 & 0.226 & B,D    \\\hline
sps7~\cite{Allanach:2002nj}  & GMSB~($\tilde{\tau}_1$ NLSP) & 0.022 & 0.016 & 0.023 & 0.015 & allowed\\\hline
sps8~\cite{Allanach:2002nj}  & GMSB~($\tilde{\chi}^0_1$ NLSP)  & 0.021 & 0.011 & 0.022 & 0.009 & allowed\\\hline
sps9~\cite{Allanach:2002nj}  & AMSB & $0.019^*$ & $0.004^*$ & $0.006^*$ & $0.002^*$ & A,B,C,D\\\hline
\hline
  SU1~\cite{Aad:2009wy} & CMSSM & 0.311 & 0.212 & 0.246 & 0.143 & D      \\\hline
  SU2~\cite{Aad:2009wy} & CMSSM & 0.009 & 0.002 & 0.010 & 0.001 & allowed\\\hline
  SU3~\cite{Aad:2009wy} & CMSSM & 0.787 & 0.440 & 0.637 & 0.258 & B,D    \\\hline
  SU4~\cite{Aad:2009wy} & CMSSM & 6.723 & 1.174 & 7.064 & 0.406 & A,B,C,D\\\hline
  SU6~\cite{Aad:2009wy} & CMSSM & 0.140 & 0.101 & 0.115 & 0.074 & allowed\\\hline
 SU8a~\cite{Aad:2009wy} & CMSSM & 0.251 & 0.174 & 0.197 & 0.120 & D      \\\hline
  SU9~\cite{Aad:2009wy} & CMSSM & 0.060 & 0.046 & 0.053 & 0.040 & allowed\\\hline
\hline
  LM0~\cite{Ball:2007zza} & CMSSM & 6.723 & 1.174 & 7.064 & 0.406 & A,B,C,D\\\hline
  LM1~\cite{Ball:2007zza} & CMSSM & 2.307 & 1.108 & 1.808 & 0.458 & A,B,C,D\\\hline
 LM2a~\cite{Ball:2007zza} & CMSSM & 0.303 & 0.201 & 0.241 & 0.139 & D      \\\hline
 LM2b~\cite{Ball:2007zza} & CMSSM & 0.260 & 0.180 & 0.205 & 0.123 & D      \\\hline
  LM3~\cite{Ball:2007zza} & CMSSM & 1.155 & 0.504 & 1.113 & 0.270 & B,C,D  \\\hline
  LM4~\cite{Ball:2007zza} & CMSSM & 0.783 & 0.432 & 0.699 & 0.260 & B,D    \\\hline
  LM5~\cite{Ball:2007zza} & CMSSM & 0.202 & 0.138 & 0.179 & 0.109 & allowed\\\hline
  LM6~\cite{Ball:2007zza} & CMSSM & 0.127 & 0.094 & 0.099 & 0.068 & allowed\\\hline
  LM7~\cite{Ball:2007zza} & CMSSM & 0.062 & 0.013 & 0.072 & 0.006 & allowed\\\hline
  LM8~\cite{Ball:2007zza} & CMSSM & 0.189 & 0.099 & 0.194 & 0.082 & allowed\\\hline
 LM9a~\cite{Ball:2007zza} & CMSSM & 0.238 & 0.029 & 0.358 & 0.015 & allowed\\\hline
 LM9b~\cite{Ball:2007zza} & CMSSM & 0.075 & 0.017 & 0.088 & 0.009 & allowed\\\hline
 LM10~\cite{Ball:2007zza} & CMSSM & 0.003 & 0.000 & 0.003 & 0.000 & allowed\\\hline
 LM11~\cite{Ball:2007zza} & CMSSM & 0.358 & 0.223 & 0.311 & 0.166 & D      \\\hline
 LM12~\cite{Ball:2007zza} & CMSSM & 0.037 & 0.008 & 0.043 & 0.004 & allowed\\\hline
 LM13~\cite{Ball:2007zza} & CMSSM & 2.523 & 0.904 & 2.289 & 0.331 & A,B,C,D\\\hline
\hline
PGM1a~\cite{Abel:2010vba} & pure GGM~($\tilde{\chi}^0_1$ NLSP)  & 0.351& 0.030&0.570&0.009& allowed\\\hline
PGM1b~\cite{Abel:2010vba} & pure GGM~($\tilde{\chi}^0_1$ NLSP)  & 0.373& 0.032&0.625&0.014& allowed\\\hline
PGM2~\cite{Abel:2010vba}  & pure GGM~($\tilde{\tau}_1$ NLSP)    & $0.008^*$& $0.005^*$&$0.009^*$&$0.003^*$& allowed\\\hline
PGM3~\cite{Abel:2010vba}  & pure GGM~($\tilde{\tau}_1,\tilde{\chi}^0_1$ co-NLSP) & 0.140& 0.103&0.121&0.086& allowed\\\hline
PGM4~\cite{Abel:2010vba}  & pure GGM~($\tilde{\tau}_1$ NLSP)  & 0.000& 0.000&0.000&0.000& allowed\\\hline
\end{tabular}}
\caption{
Status of SUSY benchmark points. For each point the columns labelled A,B,C and D give the cross section for each of the signal regions used in the ATLAS analysis~\cite{daCosta:2011qk}. The last column shows which of the four regions the point is excluded by using the new data. In the GMSB scenerio the NLSP was taken to be stable on collider time scales.
The starred cross sections are computed at leading-order values whereas all the
other values are NLO.}
\label{tab:benchmark}
\end{center}
\end{table}

\newpage

\section{Conclusions}\label{conclusions}
In this paper we have obtained the first constraints on gauge mediation models
of SUSY breaking 
with long-lived NLSP from 35pb$^{-1}$ of LHC data at 7 TeV involving jets and
missing energy in the final states. 
In carrying this out we performed an independent Monte Carlo simulation
of signal events implementing all experimental constraints
imposed by ATLAS in~\cite{daCosta:2011qk}.

\begin{figure}[!!h]
\begin{center}
\includegraphics[width=16cm]{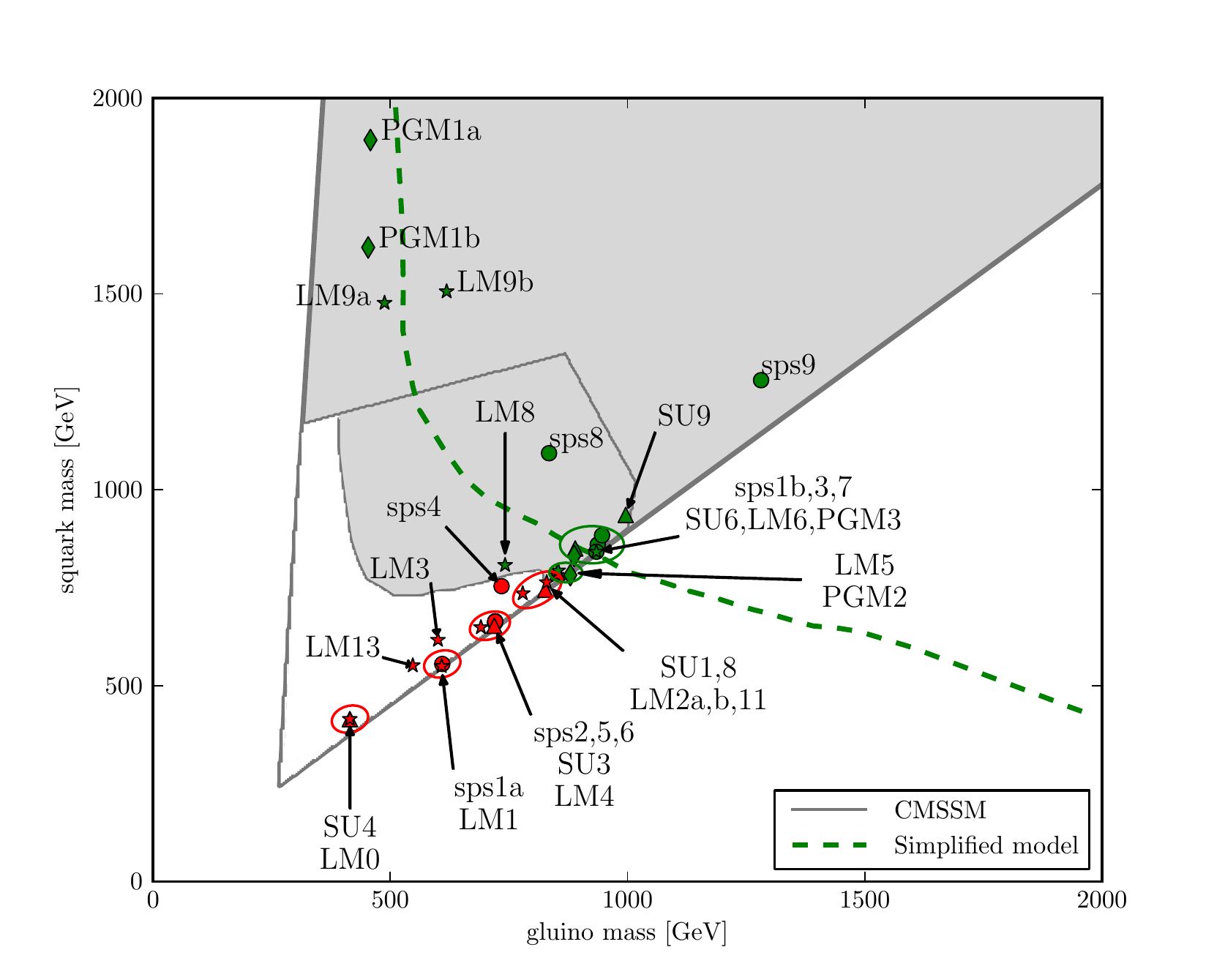}
\caption{This plot shows constraints on the CMSSM for $\tan\beta=3$, $A_0=0$ and $\mu>0$ mapped into the plane of the physical squark (average of first generation) and gluino
 masses. The kite-shaped area shows the same region of parameter space as in Fig.~\ref{fig:ATLAS-CMSSM}. The grey area is still allowed, whereas the white region inside the kite
is now excluded by the ATLAS measurements~\protect\cite{daCosta:2011qk}. The region below the diagonal
$m_{\tilde{g}}\lesssim m_{\tilde{q}}$ is not part of the CMSSM parameter space due to the influence of the
gluino mass on the squark masses during the RG evolution.
The dashed green line
gives the constraints obtained from a simplified model (containing only squarks and gluinos and a massless neutralino) 
in~\cite{daCosta:2011qk}. The reduced sensitivity in the CMSSM is mainly due to the non-negligible neutralino mass. The labelled points are the benchmark points of Tab.~\ref{tab:benchmark}. Red points are now excluded whereas green points are still viable.
}
\label{masses1}
\end{center}
\end{figure}
\begin{figure}[!!h]
\begin{center}
\vspace{-1cm}
\subfigure{
\includegraphics[width=7.5cm]{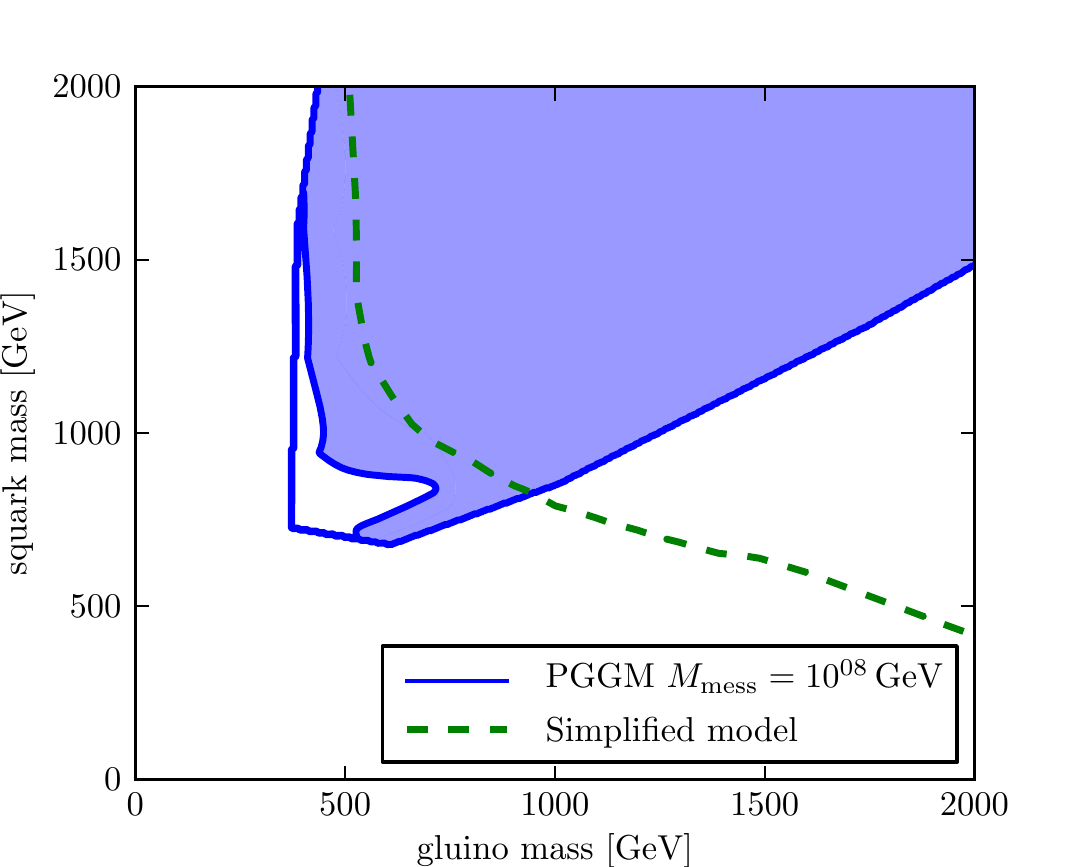}
}\\[-0.55cm]
\subfigure{
\includegraphics[width=7.5cm]{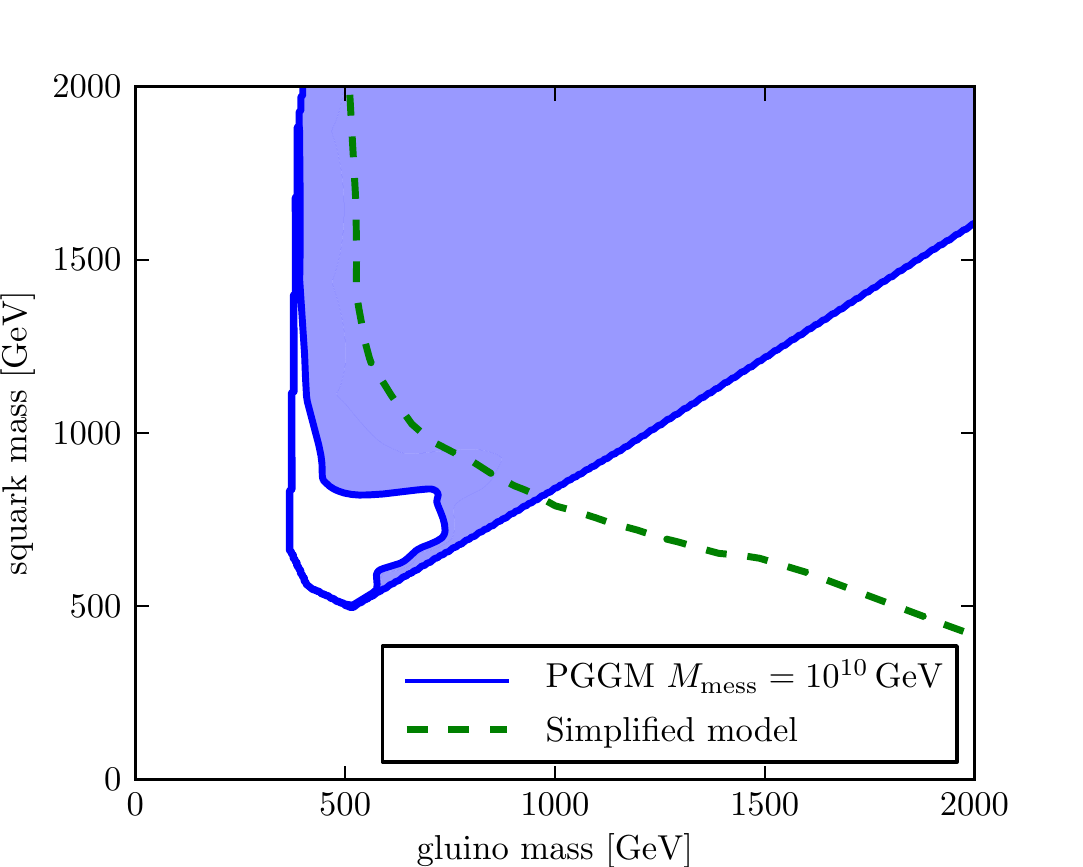}
}\\[-0.55cm]
\subfigure{
\includegraphics[width=7.5cm]{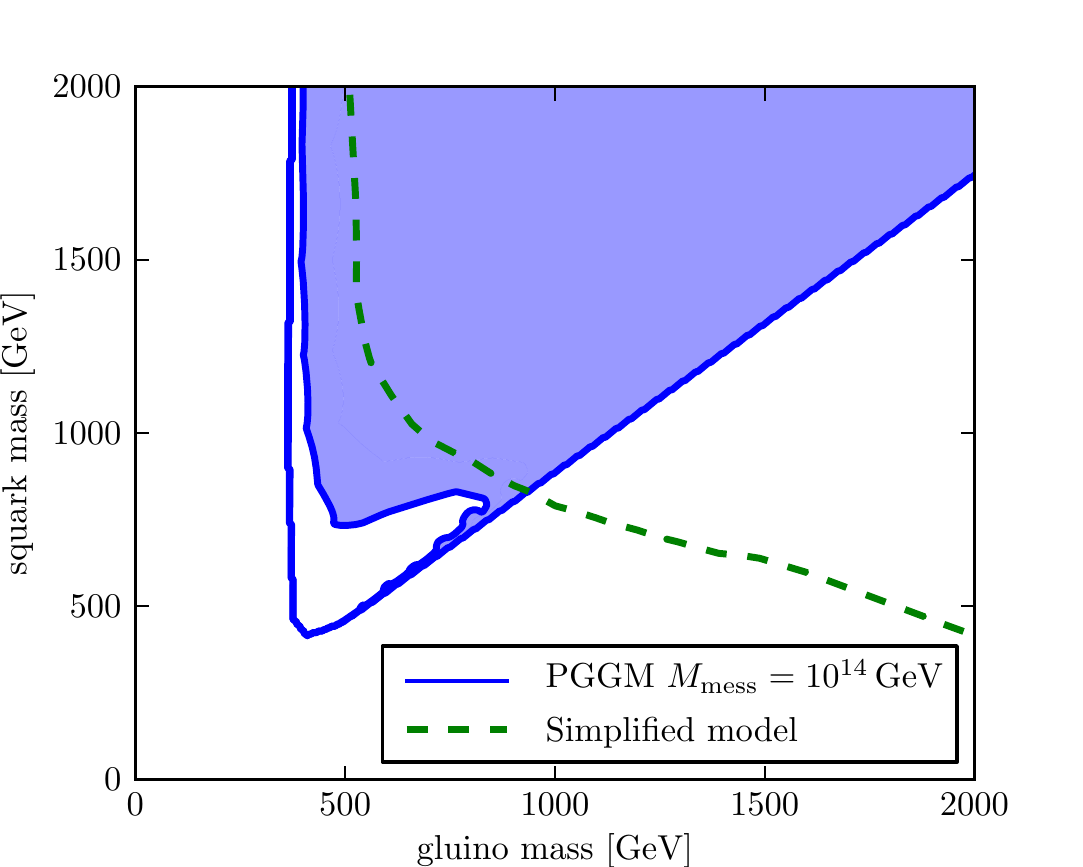}
}
\caption{Constraints on pure GGM in the plane of the physical squark and gluino masses
for three different choices of the messenger scale. For each plot the wedge-shaped region shows the previously allowed parameter space, the white part of which is now excluded by the ATLAS results. As before, the green dashed line
gives the constraints for a simplified model. The allowed region near the lower boundary 
of the model space is due to lack of missing energy in models with a stau NLSP.
}
\label{masses2}
\end{center}
\end{figure}
\begin{figure}[!!h]
\begin{center}
\includegraphics[width=16cm]{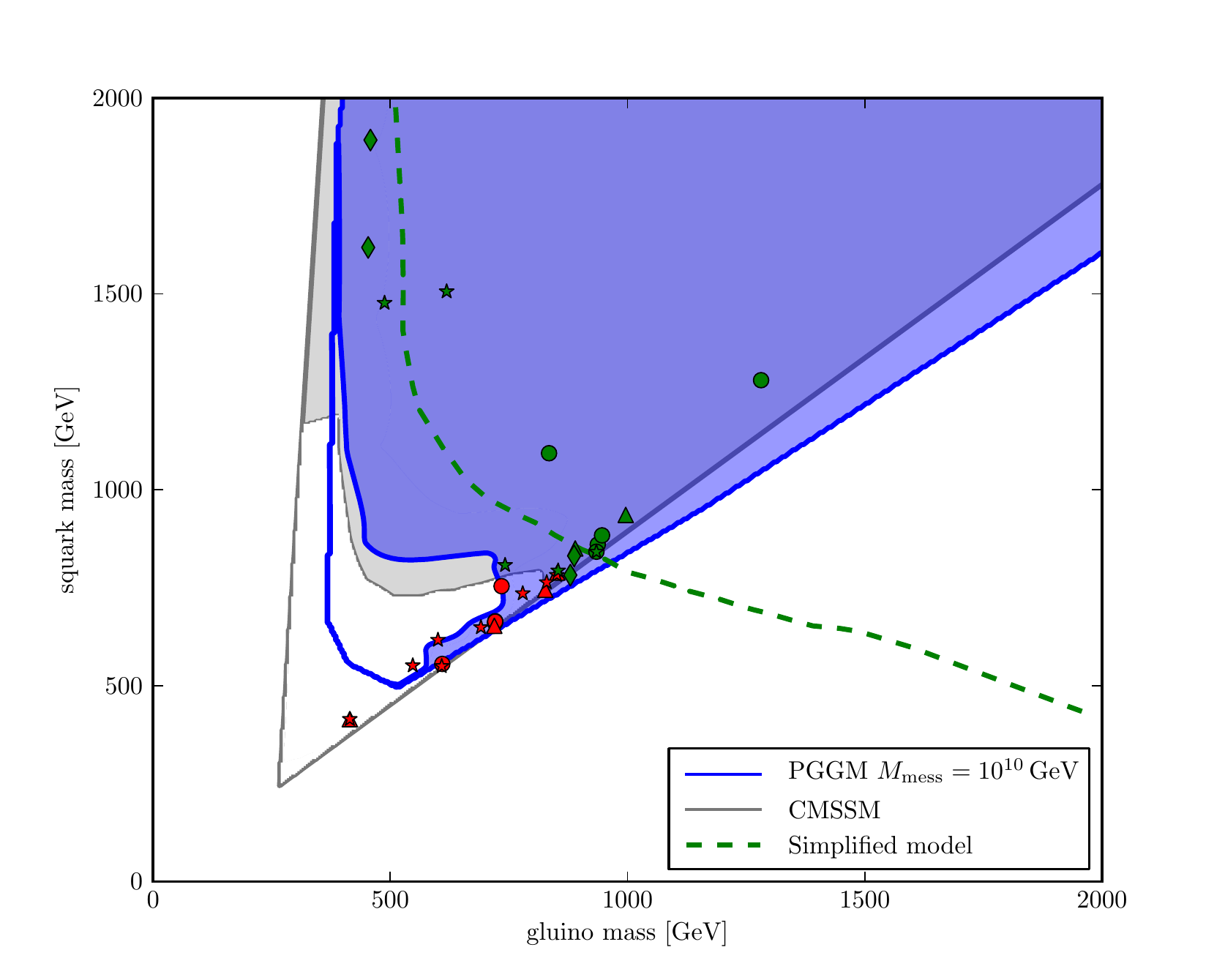}
\caption{Compilation of constraints in terms of physical gluino and squark masses in different models of SUSY breaking. In blue we show constraints in a pure GGM model with $M_{\rm{mess}} = 10^{10}$~GeV. The white region enclosed in blue lines is now excluded by the LHC data~\cite{daCosta:2011qk}, the shaded area is still viable.
For comparsion we show in grey the allowed and excluded regions for the CMSSM~(with $\tan\beta=3$, $A_0=0$ and $\mu>0$). 
We also show benchmark points from~\cite{Aad:2009wy,Allanach:2002nj,Ball:2007zza,Abel:2010vba}~(see Tab.~\protect\ref{tab:benchmark}). The green points are still allowed, and the red ones are now
excluded.
The dashed green line
 gives the constraints obtained from a simplified model as before.
}
\label{masses3}
\end{center}
\end{figure}

Our main results are summarized in Fig.~\ref{nessie} and~\ref{masses2} for gauge mediation and in Tab.~\ref{tab:benchmark} and Figs.~\ref{masses1} and~\ref{masses3}
for a standard set of benchmark points including also non-gauge-mediated models and the CMSSM.

  In addition to an interpretation of their results in terms of the
CMSSM in \cite{daCosta:2011qk} ATLAS also presented their results as a limit on a squark
and gluino masses in a simplifed model which only contained squarks, gluinos and
a massless neutralino. The latter tends to increase the search reach by increasing the
energy of the visible decay products.

In Figs.~\ref{masses1}-\ref{masses3} we interpret and combine the ATLAS constraints on different SUSY models by displaying them in 
terms  of physical squark and gluino masses. Fig.\,\ref{masses1} shows the mapping
of the CMSSM $m_0$-$m_{1/2}$ parameter space into the squark-gluino mass space. We note that the region
$m_{\tilde{g}}\lesssim m_{\tilde{q}}$ is not accessible in any of the models we have studied due
to the effects of the gluino masses on the squark masses during the RG evolution\footnote{While close to one in the CMSSM, 
the slope of the boundary in $(m_{\tilde{g}},m_{\tilde{q}})$ is model dependent.}.
In general, constraints in the squark and gluino plane are similar in all the models we have considered~(see Figs.\,\ref{masses1}-\ref{masses3}), 
despite significant differences in the mass spectra of the other sparticles.
Importantly however this can change if the (N)LSP is not the lightest neutralino
as can be seen in Fig.\,\ref{masses2}. Moreover this limit is always weaker than that obtained in 
the simplified model due to the non-vanishing neutralino mass.

\newpage

\newpage
\enlargethispage{2\baselineskip}

\newpage

\newpage

\section*{Acknowledgements}
We are grateful to Steve Abel, Alan Barr, Oliver Buchmueller, T. J. Khoo and 
Chris Lester for interesting discussions and comments. This work was supported
by STFC.

\providecommand{\href}[2]{#2}\begingroup\raggedright\endgroup
\end{document}